# Effect of surface morphology on magnetization dynamics of cobalt ultrathin films: An in-situ investigation

Anup Kumar Bera[1], Pooja Gupta[2], Debi Garai[3] and Ajay Gupta[3] Dileep Kumar[1*],

[1]UGC-DAE Consortium for Scientific Research, Khandwa Road, Indore-452001, India
[2] Synchrotron Utilization Division, RRCAT, Indore 452013, India
[3]Amity Centre for Spintronic Materials, Amity University, Sector 125, Noida-201313, India
[*]Email: dkumar@csr.res.in

**Abstract**

Growth of Co film on $SiO_2$ substrates with surface roughness of 0.5 nm and 1.6 nm have been studied in-situ using magneto-optical Kerr effect (MOKE) and four probe resistivity measurements. In-situ measurements jointly suggests that the films grow via Volmer-Weber growth process and proceed via a nonmagnetic, superparamagnetic and a ferromagnetic phase formation on both the substrates. Islands are found to coalesce at film thicknesses ~ 0.6 nm and at ~ 1.5 nm with continuous film formation around film thickness of ~1.5 nm and ~ 3.0 nm for smooth and rough substrate respectively. Ferromagnetic long-range ordering i.e., appearance of magnetic hysteresis loop in both films is observed just after coalesce stage. Observed azimuthal angular dependence of coercivity confirmed the presence of a weak uniaxial magnetic anisotropy (UMA) in both the films, whereas difference in UMA with substrate roughness is interpreted in terms of combined effect of domain wall pinning and internal stresses in the films. Origin of much higher UMA in case of the Co film deposited on ripple patterned substrate of similar root mean square roughness, is attributed to the modified long range dipolar stray fields on the surface.

## INTRODUCTION

Thin film magnetism is an important area of condensed matter physics for its wide variety of application in the area of spintronics [1], magnetic sensors [2], magnetic memory [3] and magneto-electronic devices [4]. In this field, surface and interface morphology at the atomic scale plays an important role which controls many important physical, chemical and magnetic properties such as surface adhesivity [5], plasmonic catalysis [6], photovoltaic chemistry [7], interlayer exchange coupling [8] , electrical conductivity [9] magnetoresistance [10] , magnetic anisotropy [11] , spin ejection efficiency [12], formation of dead layer [13] etc. that dictates the performances of many thin film-based devices . Apart from this, in basic elementary studies of quantization and interaction phenomenon arising due to reduction of element size increases the importance to explore the surface roughness effects. Therefore, an interest has been growing to study the effect of surface and interface morphology on different branches of experimental physics.

Magnetism is a collective phenomenon and strongly sensitive to local environment. Therefore, magnetic properties such as domain structure, magnetization reversal, coercivity, hysteresis, magnetoresistance are strongly affected by surface as well as interface morphology [14][15]. Even most studies done in ultrathin films of current interest are grown over some substrate that are not absolutely flat and smooth. Rather, it contains oriented steps and finite roughness. Studies have found that it contributes an extra magnetostatic energy contribution to the overall energy of the system [16]. Moreover, as we go away from idealized surface and deal with nanostructured film with complex profile, the situation gets complicated due to interplay of different effects [17].Therefore, more knowledge on intercorrelation and separation of different contribution between the morphology with growth, transport and magnetic properties are required. Moreover, a systematic modification of roughness and morphology is major requirement for better realization of underlying effects. In the view of the above fact, magnetic thin films and multilayer structures have been extensively studied in literature [17][18][19][20][21] . However, investigations of this kind are still in the progress. It may be noted that most of such studies were either performed *ex-situ* or in embedded systems, where the thin magnetic films are covered with a protective nonmagnetic layer. However, in both the cases, the oxidation of the surface or nonmagnetic protective layer can significantly modify the magnetic as well as transport properties . Therefore, it is difficult to study unambiguous effect of roughness on peculiar magnetic properties. *In-situ* measurements under ultra-high vacuum (UHV) condition is more useful to obtain a genuine property of thin magnetic layer itself without an over layer[11][14][22][23]. Furthermore, it also provides detailed thickness dependence information from same sample [18][24]

In the present work, the effect of substrate surface roughness on the growth of Co film with thickness ranging from fraction of nanometer to several nanometer, has been studied using combined *in-situ* magneto-optical Kerr effect *(*MOKE) and four probe resistivity measurements. In-situ investigation in the present case not only provided magnetic properties of the thin film without contamination but also the combined analysis of their magnetic, transport and structural properties provided a complete picture of Co film growth on $SiO_2$ substrates with different surface roughness. Co films are also grown on an ion beam sputtered ripple patterned substrate and results are discussed in context of randomly corrugated substrate morphology.

# EXPERIMENTAL SECTION

Cobalt films of thickness ~15 nm was grown on $SiO_2$ substrates with root mean square (rms) roughness $\sigma$ = 0.50 nm and 1.6 nm (designated as $GL\sigma_{0.5nm}$ and $GL\sigma_{1.6nm}$ respectively) using electron beam evaporation technique inside an UHV chamber having a base pressure $5\times10^{-10}$ mbar or better. Chamber is equipped with facilities for film characterization techniques such as MOKE, four-probe resistivity, X-ray reflectivity (XRR) and reflection high energy electron diffraction (RHEED) for in-situ investigation of magnetic, transport and structural properties. Growth of Co film on both the substrates were investigated by performing in-situ four probe and MOKE measurements simultaneously during the film growth. For both the films, in-situ measurements were done one by one under identical condition (same vacuum conditions and identical growth parameters) in order to study roughness effect unambiguously. Surface morphology of both the films were imaged by means of AFM in contact mode. Grazing incidence XRR measurements were also performed using Bruker D8 diffractometer for accurate extraction of film roughness and thickness. Ex-situ MOKE measurements were also performed with the external magnetic field (H) applied along different directions in the plane of the film in order to get information about magnetic anisotropy in the films. In order to draw a comparative study on film morphology, crystallinity and magnetic anisotropy with an ordered surface roughness, a periodic ripple patterned $SiO_2$ substrate is prepared by oblique Ar+ ion beam sputtering of energy 700 eV and average ion flux ~ $1.3\times 10^{14}$ ions/$cm^2$ sec. During ion irradiation on the substrate the angle of ion incidence was kept fixed at 65º from the surface normal. Co film of nominal thickness 35nm was deposited on this substrate in the same UHV chamber and its crystallinity, morphological and magnetic characterization was performed by RHEED, AFM and MOKE measurement respectively.

# RESULTS AND DISCUSSIONS

Fig. 1a represents the XRR patterns of the $SiO_2$ substrates obtained prior to the in-situ measurements. Surface roughness of $\sigma$= 0.5 nm and 1.6 nm were obtained by fitting the experimental XRR data using Parratt's formalism [25]. In order to get information about the in-plane roughness, X-ray diffuse scattering measurements are done by keeping the scattering angle $\theta$ fixed and varying the angle of incidence from 0º to $2\theta$. Measurements are reported for $\theta=0.7º$. Fig. 1b gives the X-ray diffuse scattering (XDS) pattern of both the samples. Some qualitative features, such as the difference in intensity between the specular and Yoneda peaks can be observed. The apparent decrease in specular peak intensity and increase in diffuse peak intensity of the $GL\sigma_{1.6nm}$ sample compared to $GL\sigma_{0.5nm}$ sample is due to higher substrate roughness. The XDS patterns have been fitted with a correlation length 80nm and 50nm for $GL\sigma_{0.5nm}$ and $GL\sigma_{1.6nm}$ samples respectively.

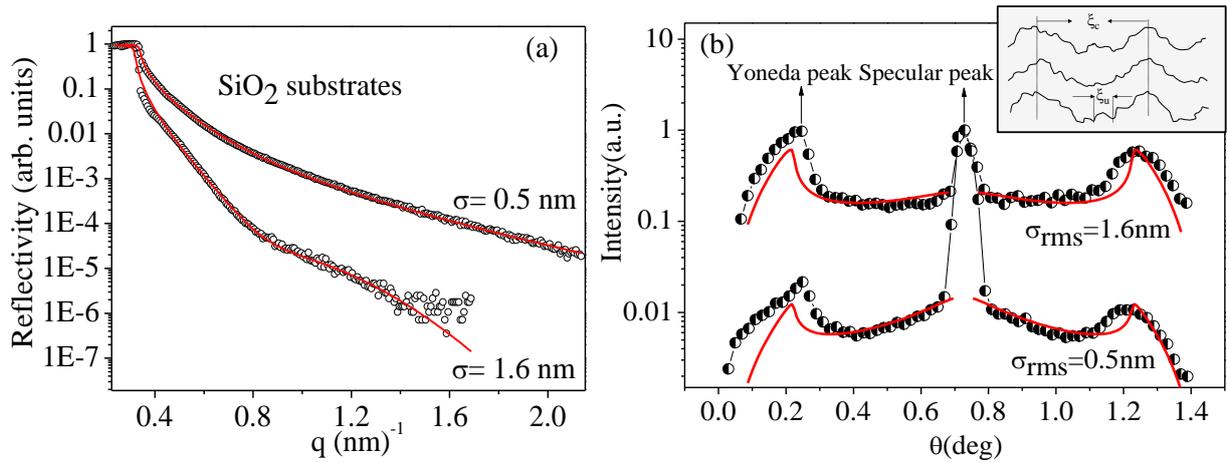

**Fig. 1.** (a) XRR patterns of the $SiO_2$ substrates. Continuous curves are the best fit to the experimental data. (b) Diffuse scattering curves for rough and smooth sample taken at θ=0.7º. Schematic in the inset shows in-plane correlation length of correlated (ξc) and uncorrelated roughness (ξu), respectively.

The evaluation of hysteresis loops with increasing Co layer thickness on both substrates is reported in Fig. 2. One may note that, in case of smooth substrate, a faint hysteresis loop starts appearing at a thickness of 0.7 nm, whereas in case of rough substrate it appears after deposition of 1.8 nm thick Co film. Therefore, it indicates that for smooth substrate long range ferromagnetic ordering sets on earlier compared to the rough substrate. This is due to the fact that more Co atoms are required to fill up the valley regions of the rough substrate for formation of continuous film. The variation of coercivity extracted from MOKE hysteresis loop is plotter in Fig. 3(a). We observe that initially Hc of the film increases with film coverage. However, after reaching a maximum value of about 27 Oe (for $GL\sigma_{0.5nm}$) around 1.5 nm thickness and 32 Oe (for $GL\sigma_{1.6nm}$) around 3.3 nm thickness, $H_C$ exhibits a slow decrease with further increment of film coverage. The absence of hysteresis loop upto Co film thickness 0.7 nm and 1.8 nm respectively for smooth and rough $SiO_2$ substrate suggests that the Co islands consist of single magnetic domain and exhibit superparamagnetism [18]. With increasing film thickness, as the islands become multi-domain, the $H_C$ starts developing. Once the film becomes continuous, $H_C$ reaches its maximum value. With further increase in the thickness, the contribution of surface irregularities which act as a pinning center for domain walls, decreases and therefore the Hc exhibits a slow decrease. Thus, the effect of substrate roughness in Co growth is well correlated with the Hc variation.

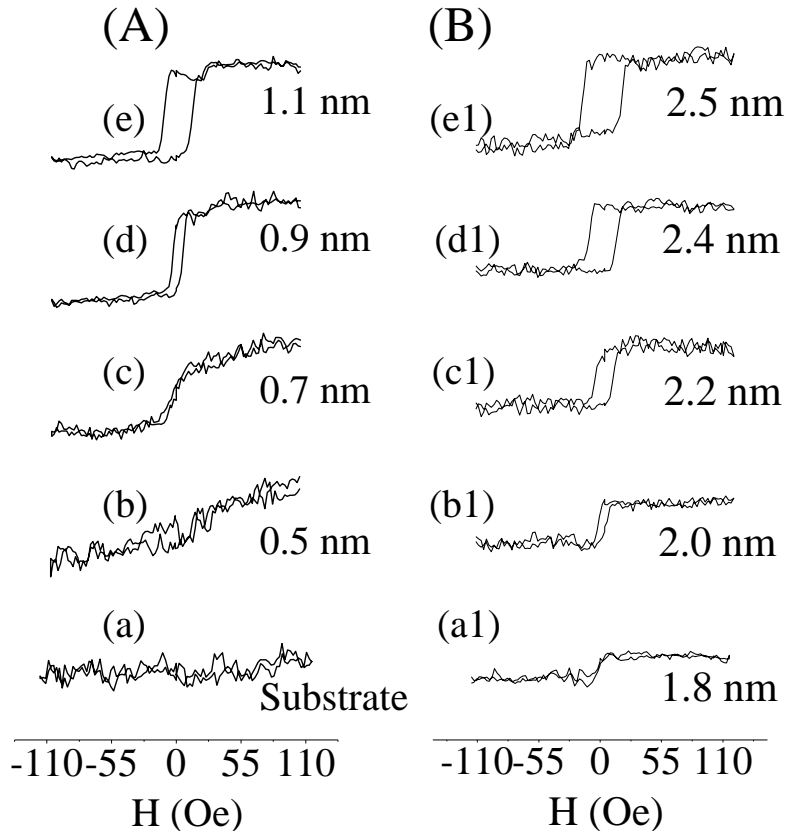

**Fig. 2.** Representative hysteresis loops with increasing Co layer thickness on [A] GL$\sigma_{0.5nm}$ and [B] GL$\sigma_{1.6nm}$ substrate.

The study of magnetic dead layer at the interface of magnetic and non-magnetic layers has been attracted considerable interest as they are always undesirable in practical applications due to their drastic degradation of host magnetism. In-situ MOKE measurement is a handy tool to study the magnetic dead layer at the interface which could be form due to some intermixing, hybridization or some structural modifications at the interface[20]. To get information about dead layer at the interfaces, thickness dependence of MOKE signal (Ms1-Ms2) is extracted from the saturation point of hysteresis loops for both the films and is reported in Fig. 3b. Due to finite penetration depth of laser beam, the linear dependence of Kerr signal for smaller film thickness can be related to the expression as given below –

$$\Phi_{long} = \frac{4\pi n_{sub} Q d \theta}{\lambda(1-n_{sub}^2)}.$$

Here Q is the magneto-optical constant, d is the magnetic film thickness, θ is the angle of incidence of laser beam from surface normal. One can see that the signal is linearly proportional to the thickness in ultra-thickness range. However, it tends towards saturation in higher film thickness due to limited penetration depth of laser. The inset in Fig. 3b shows the enlarged curve of linear fit in the lower thickness region.

The extrapolated lines cut the x axis at almost near zero thickness in both cases. This suggests that there is no magnetic dead layer at the interfaces of both substrates. In general, an amorphous oxide layer prevents Co to reach in the proximity of Si and possible formation of magnetically dead cobalt silicide layer[20]. Therefore, in the present case no dead layer is formed at the substrate and Co interface. It may be noted that for smooth substrate, the MOKE signal vs film thickness graph has stepper slope compared the rough one. This is due to larger dipolar field generated by rougher film surface that partially reduces ferromagnetic ordering and hence MOKE signal.

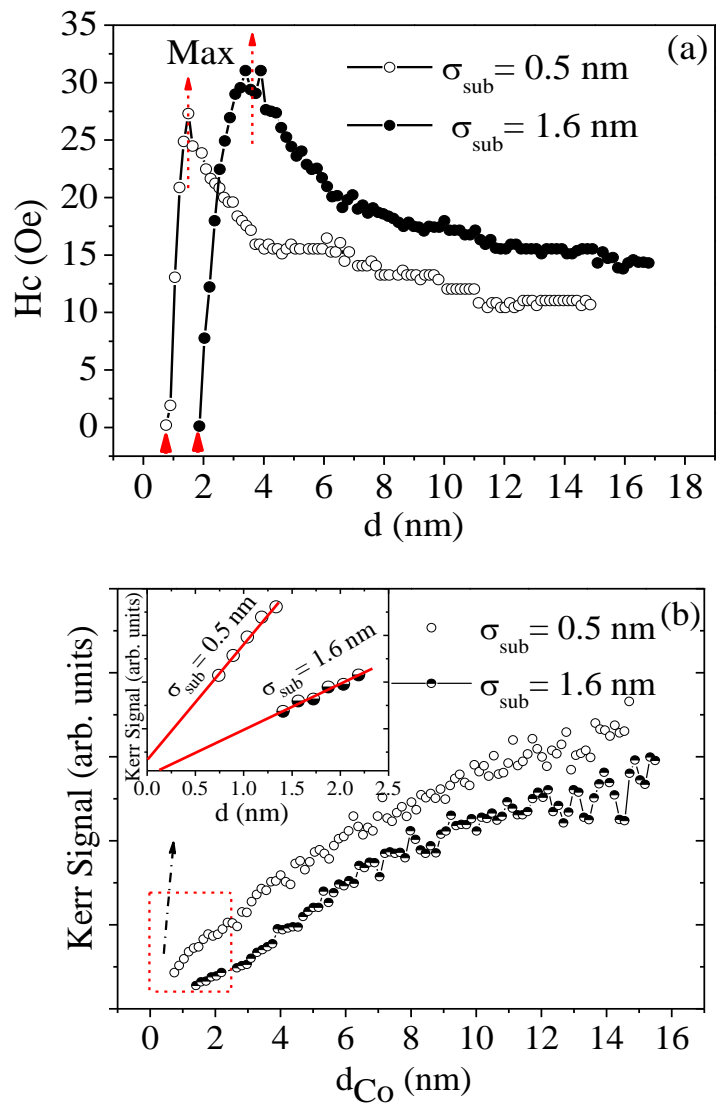

**Fig. 3.** Variation of (a) $H_C$ (b) and Kerr signal as a function of Co growth on rough and smooth substrate.

The schematic of the four-probe resistivity measurement used for the present study is shown in Fig 4a. Fig. 4b gives the thickness dependence of sheet resistance ($R_{Co}$) obtained through simultaneous four probe resistivity measurements. $R_{Co}$ remains almost constant upto a thickness about 0.5 nm for $GL\sigma_{0.5nm}$ and about 1.2 nm for $GL\sigma_{1.6nm}$ substrate. It confirms that Co films grow via Volmer-Weber type of growth on both substrates which is also expected due to large difference in surface free energy between Co and $SiO_2$ [20]. Furthermore, this behaviour could be understood considering the fact that up to these thicknesses, conduction of electrons from one island to another will not take place due to the separation between Co islands. With further increase in the thickness, a rapid decrease in the $R_{Co}$ for both the films confirm the formation of percolating path around a thickness of about 0.6 nm and 1.4nm for smooth and rough substrates, respectively. These percolating thickness values are estimated by differentiating the $R_{Co}$ vs film thickness plot as shown in the inset of Fig. 4b. It may be noted that in the thickness range of about 1.5–2.0 nm a continuous film is formed on smooth substrate whereas it is about 3.0–3.5 nm in case of rough substrate. Beyond these thicknesses, $R_{Co}$ exhibit a slow decrease because of decreasing contribution of surface and interface scattering. Therefore, a combined analysis of thickness dependent $R_{Co}$ and $H_C$ data for both substrates are used to understand growth behavior of the Co film on $SiO_2$ substrates and the same has also been correlated with the substrate surface roughness.

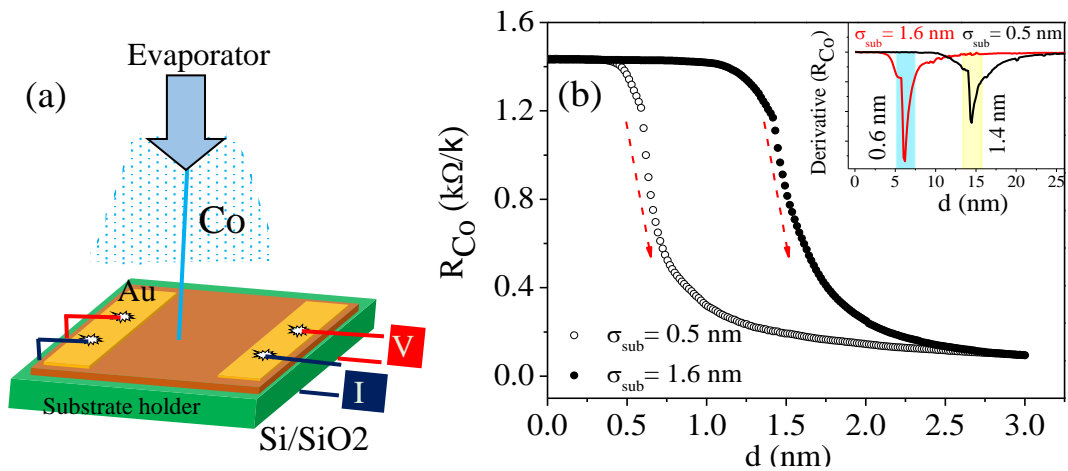

**Fig. 4.** (a) Schematic of four probe sheet resistance measurement. (b) Variation of sheet resistance as a function of Co film thickness(d) on rough and smooth substrate ($GL\sigma_{0.5nm}$ and $GL\sigma_{1.6nm}$). The inset shows the derivative of the $R_{Co}$ vs d plot obtained for estimation of percolating film thickness

After performing in-situ MOKE and resistivity experiments on both samples, the samples are taken out from the UHV chamber and are studied ex-situ using XRR, AFM and longitudinal MOKE measurement. X-ray reflectivity is used to get the thickness and rms roughness of the Co films. The information about a possible magnetic anisotropy is obtained by performing the longitudinal MOKE measurement in the different in-plane direction of the samples.

The XRR patterns of both films are presented in Fig. 5. The Kiessig fringes seen in the reflectivity spectra come from the constructive interference between X-rays reflected from film/air interface and film/substrate interface. We observe that the slope of the XRR pattern of Co/$GL\sigma_{1.6nm}$ sample is larger compared to the Co/GL$\sigma_{0.5nm}$ sample. This is due to the large x-ray diffuse scattering from the rougher surface. XRR data of the both films fitted using Parratt's formalism and the fitting parameters of the XRR pattern are given in Table 1. In order to get good theoretical fit of the data, it was found necessary to incorporate a thin oxide layer with lower electron density. This surface layer has an electron density ~12% less than that of the film and may be attributed to surface oxidation.

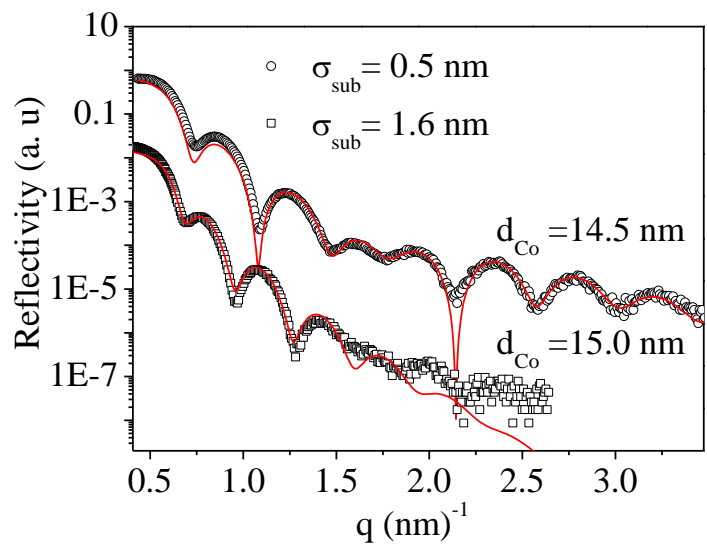

**Fig. 5.** X-ray reflectivity pattern of the Co films on GL$\sigma_{0.5nm}$ and GL$\sigma_{1.6nm}$ substrates. Continuous curve represents the fitting of the experimental data using Parratt's formalism.

AFM measurements have also been performed for both the samples in order to get more information about the roughness and morphology of Co films. The corresponding AFM images are shown in figure 6a and 6b respectively. The surface morphology is completely isotropic and consists full of small islands or mounds and pits. In order to get the statistical roughness parameters, such as surface roughness and lateral correlation length ($\xi$) of both the films, information about the height-height correlation function (H(r)) has been obtained from image analysis of a 2μm×2μm AFM image. Height -height correlation function is defined as H(r)=[<(Z(r)-Z(0))$^2$>$_{area}$]. Where Z(r) and Z(0) are the height at the coordinates at r=(x,y) and (0,0), respectively. The bracket in the above expression is spatial average. Therefore, from the AFM images H (r)Vs R(r) were directly extracted by averaging over several distinct regions of large extent [26] and shown in Fig. 6c and d. As the height- height correlation function can also be express as H(r) = 2$\sigma^2$ [1-exp (-(r/$\xi$))$^{2\alpha}$], therefore, correlation length ($\xi$) including other roughness parameter $\sigma$ (rms roughness), roughness exponent ($\alpha$) are obtained by fitting experimental plots (figure 6c and d) using this equation [26]. The obtained values of $\xi$ are presented in Table 1.

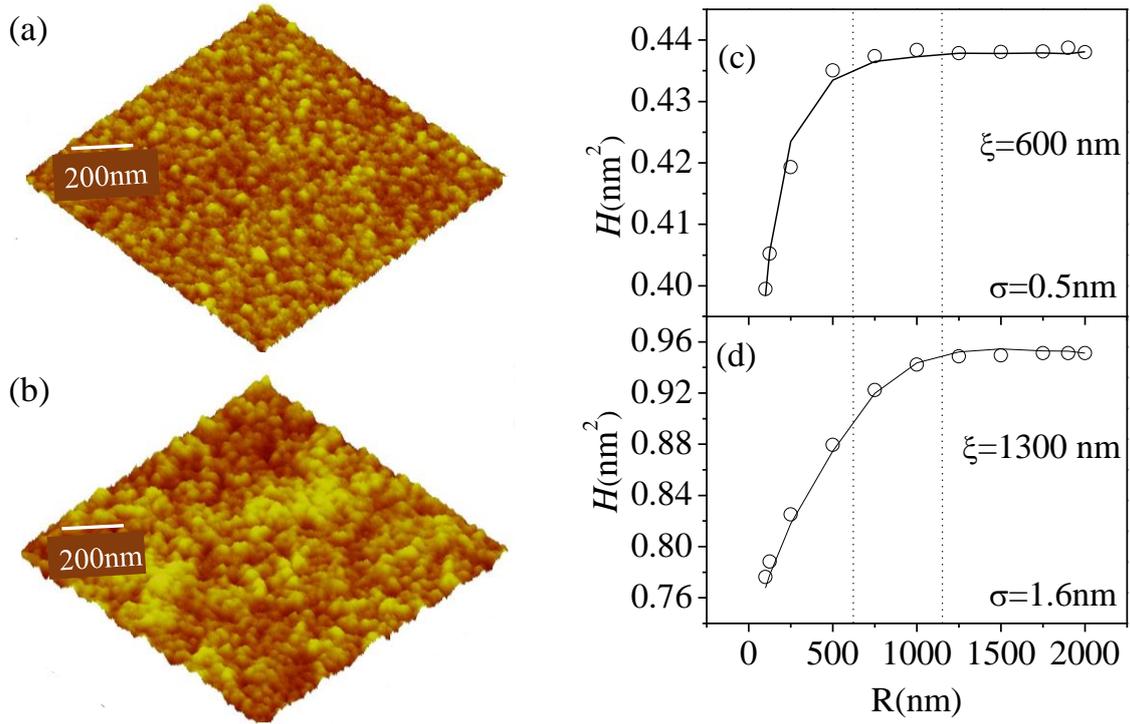

**Fig. 6**. AFM images of Co film on (a) $G_{\sigma=0.5nm}$ and (b) $GL_{\sigma=1.6nm}$ substrates. (c) & (d) are corresponding height-height correlation function, H (r), extracted from AFM images, respectively. Continuous line gives experimental data fitted with the help of theoretical model.

**Table1:** The fitted parameters of XRR data and height-height correlation function (H(r) function obtained from AFM images). $d_{Co}$ is the thickness of the Co film; $\sigma_{sub}$ and $\sigma_{sur}$ are the r.m.s. roughness of bare substrate and film respectively; $\sigma_{AFM}$, $\xi$, are the roughness parameters obtained from the AFM data analysis and N is the demagnetization factor.

| Sample | Using XRR | | | Using AFM | | |
|---|---|---|---|---|---|---|
| | $\sigma_{sub}$ (nm) | $d_{Co}$ (nm) | $\sigma_{sur}$ (nm) | $\sigma_{AFM}$ (nm) | $\xi$ (nm) | $N(10^{-4})$ |
| Smooth | 0.5 | 14.5 | 0.60 | 0.43 | 600 | 0.94 |
| Rough | 1.6 | 15.0 | 1.30 | 0.95 | 1300 | 1.26 |

From Table 1, it is interesting to note that in sample Co/GL$\sigma_{0.5nm}$, interface roughness remains comparable to the substrate roughness, whereas in sample Co/GL$\sigma_{1.6nm}$, the interface roughness is lower than that of the substrate roughness. Thus, in case of highly rough surface the deposition of ~ 15 nm Co layer has some smoothening effect. The results are in accordance with some earlier studies [27]. Roughness obtained using XRR are qualitatively in good agreement with AFM results but quantitively the results are different. The rms roughness measured by XRR is usually larger than those by AFM. This is due to the finite sized AFM tip smears out high frequency contributions to the surface profile and is unable to follow the surface profiles into the deep but narrow valleys. It may also be noted that substantially higher $\xi$ (1300nm) for rough substrate is as per expectation, since higher roughness substrate is having bigger hills and valley like structure in the surface morphology. To understand the effect of roughness on magnetic properties, the surface demagnetization (N) is calculated by Schlömann equation[28] -

$$N = \pi\sigma^2/\xi d. \tag{2}$$

Where d is the film thickness, $\sigma$ is rms roughness, and $\xi$ is the lateral correlation length. The calculated N for both samples is given in the Table 1.

To get information about the magnetic anisotropy in both the samples, hysteresis loops were measured as a function of in-plane azimuthal angle ($\varphi$) ranging from $\varphi=0º$ to $\varphi=360º$. The representative hysteresis loops at $\varphi=0º$ and $\varphi=90º$ for Co/GL$\sigma_{0.5nm}$ and Co/GL$\sigma_{1.6nm}$ samples are given in the Fig. 7a and b, respectively. There is a strong variation in the shape of the hysteresis curve with azimuthal angle. For $\varphi=0º$ the hysteresis curve is a perfect square loop suggesting that the magnetization reversal takes place purely through domain wall motion. With increasing azimuthal angle, the rounding-off of the hysteresis curve indicates increasing contribution of the rotation of domain magnetization. Therefore, $\varphi=0º$ & $90^0$ direction corresponds to the easy and hard axis of the in-plane magnetic anisotropy present in both samples. The observed angular dependence of the coercivity determined from the corresponding MOKE measurements is plotted in Fig. 7(c) and (d) respectively. It may be noted that the value of Hc in all angular range is larger for sample Co/GL$_{\sigma1.6nm}$ as compared to Co/GL$_{\sigma0.5nm}$, which indicates that Hc has increased with the surface roughness. The above results indicate that the easy magnetization direction of the Co films lie in the film plane for both the samples and a weak uniaxial magnetic anisotropy is also present.

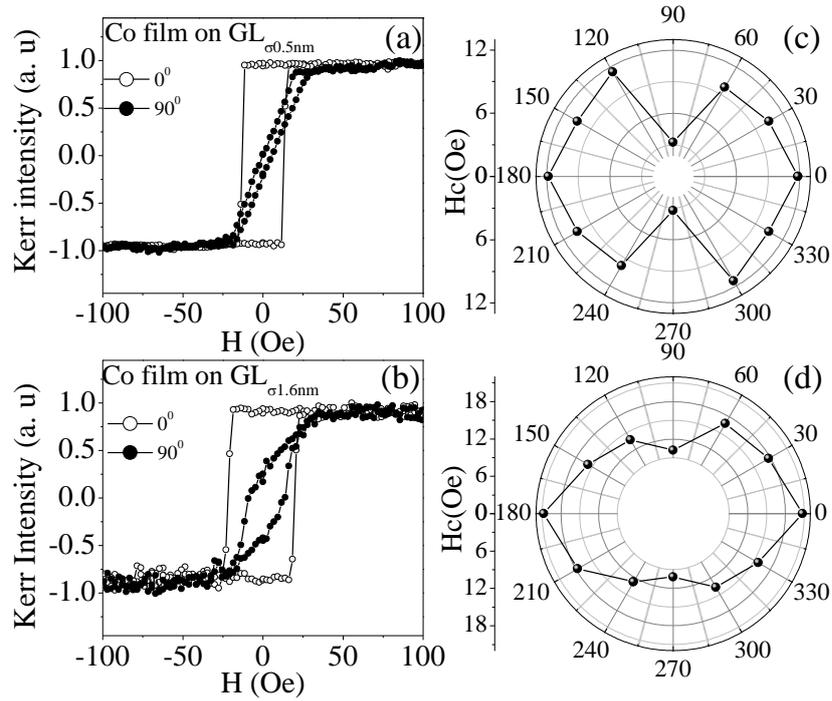

**Fig.7.** The hysteresis loop along easy direction($\varphi=0°$) and hard direction ($\varphi=90°$) for (a) Co/GL$\sigma_{0.5nm}$ and (b) Co/GL$\sigma_{1.6nm}$ samples along with azimuthal angle dependence of coercivity (polar patterns) (c and d) are shown.

To develop further understanding the role of surface roughness on film morphology, its crystallinity and magnetic anisotropy we have extended this study for a periodically rough substrate. For this purpose, Co film is also grown directly on ripple patterned SiO$_2$ substrate having average periodicity of 65nm and rms roughness 1.87 nm. It may be noted that rms roughness of this substrate is nearly equal to the GL$\sigma_{1.6nm}$ substrate, but the surface morphology is in the form of periodic ripples. AFM image and fitted XRR pattern of the rippled substrate taken prior to the film deposition is shown in Fig. 8 (a) and 8(b), respectively.

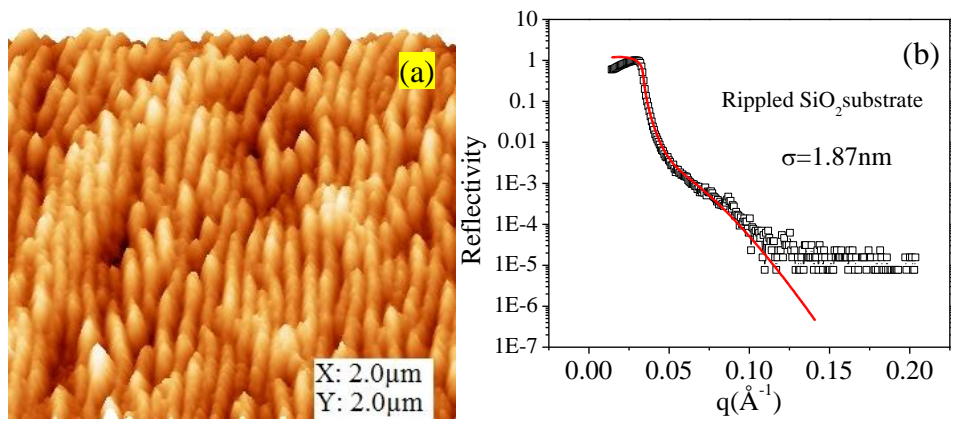

**Fig. 8.** (a) AFM image and (b) XRR pattern of the rippled patterend SiO$_2$ substrate.

Co film of nominal thickness 35 nm as readout from a calibrated quartz crystal monitor is deposited over the rippled substrate inside the same UHV chamber. After deposition, film crystallinity was checked by RHEED measurement. The in-situ RHEED pattern of the film is given in Fig. 9(a). It consists of concentric Debye rings with uniform intensity distribution. Therefore, it indicates that the film grows in polycrystalline state without any texturing. The rings corresponding to different planes of hexagonal closed pack Co structure are indexed in the image. The morphology of the film imaged by AFM is reported in Fig. 9(b). Clear traces of periodic ripples are easily visible in the image. From this AFM image, we obtain an average wavelength λ= 64nm and $\sigma_{rms}$=2.84nm of the film surface. The presence of periodic ripple pattern of same wavelength as that of substrate indicates that the film replicates the substrate morphology.

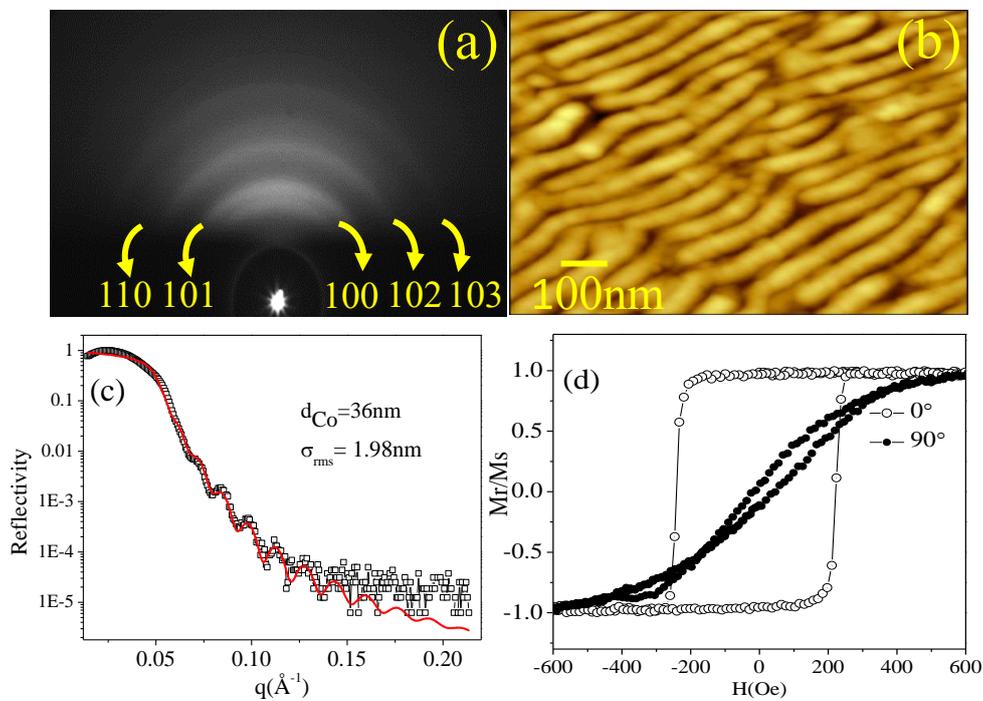

**Fig. 9.** (a) RHEED pattern and (b) AFM image of Co film grown on rippled $SiO_2$ substrate. The corresponding (c) XRR pattern and (d) MOKE hysteresis loop taken along and normal to the ripple direction.

The corresponding reflectivity pattern of the film is reported in Fig. 9(c). The data is fitted with a film thickness 36nm and rms roughness 1.98nm using Parratt formalism. In order to obtain magnetization reversal behaviour, MOKE hysteresis loops have been taken by applying external magnetic field parallel to different in plane directions. The MOKE hysteresis loop taken parallel and perpendicular to the ripple direction is shown in figure 9(d). We found that the hysteresis loop taken parallel to ripple direction displays rectangular shape loop whereas perpendicular to ripple direction it displays almost reversible behaviour with negligible opening at the centre. The film has a coercivity of 243Oe along easy axis whereas, the hard axis saturation field is about 680Oe which is one order magnitude higher compared to previous two films. Thus, the film exhibits a pronounced UMA. The presence of easy axis parallel to the ripple direction indicates that the origin of UMA is strictly related to the shapes of the ripple. Therefore, the main source of UMA in this case is periodic modulation of film topography that creates dipolar stray fields or shape anisotropy [11][14][29][30]. Thus, from the above observation we find that the rippled patterned morphology does not have any influence on the crystallinity of the film. However, it induces a strong uniaxial magnetic anisotropy solely originated from shape anisotropy compared to the randomly oriented substrate roughness. Thus, to achieve a desired strength of UMA, systematic studies on the dependencies of UMA over ripple wavelength, roughness, film thickness is worthwhile.

Combined in-situ experiments and analysis revealed that surface morphology of the substrate modify magnetic and transport properties significantly. In case of the higher rough substrate, increased coercivity throughout the full range of thickness can be understood in terms of in-plane demagnetization factor caused by magnetic poles. It is found that the film with higher roughness has higher demagnetization factor. If the in-plane demagnetization factor increases, in order to achieve the same magnetization field inside the material, one needs to increase the applied field. Immediately, one can connect this with coercivity measurement of rough Co thin film, which increases with demagnetization factor. According to the study of Zhao et al. [16], the in-plane coercivity changes linearly with the in-plane demagnetization factor. The general trends for magnetic thin film are that the coercivity increases with film roughness[31][32], which seems to be agreed with our experimental results.

Origin of magnetic anisotropy and its correlation with surface morphology is important to get polycrystalline thin films with desired magnetic properties. It is important to note that due to random orientation of grains in polycrystalline Co films, magneto-crystalline anisotropy is expected to be absent. Moreover, surface morphology of the films is also isotropic and random. Therefore, magnetocrystalline anisotropy and dipolar interaction both cannot be responsible for the in-plane preferential alignment of the spins. Thus, considering the above facts, the presence of sizeable amount of UMA in the present case may be attributed to a long-range stress which might have developed during film deposition [33] [34][35]. In case of the rough substrate, decrease in UMA can be understood in terms of increase in random stress in the film. The long-range stresses give rise to the uniaxial anisotropy due to magnetoelastic coupling [36], whereas short-range stresses are random in nature [37]. In case of the rough Co surface, the short-range stresses and domain pinning start contributing against UMA. In view of the fact, the effective UMA, which is defined by the long-range stresses in the film decreases, and the field needed to switch magnetization reversal increases. In case of film deposited on rippled substrate, the long ranged dipolar interaction between the well-organized ripple crests creates stronger shape anisotropy. Moreover, the external magnetic field must overcome the pinning energy as well as the component of magnetic anisotropy parallel to applied field direction for magnetization reversal. Therefore, both the induced UMA and coercivity are increased significantly. Extensive research work is being conducted, where the contribution of the stray dipolar fields, generated due to the asymmetric morphology (rippled geometry) of the film surface, is being used to induce UMA in the polycrystalline thin films and multilayered structures[33][38][39][40] .

## CONCLUSION

Growth and magnetic hysteresis behaviour of Co film grown over $SiO_2$ substrate of different surface morphology has been studied during in-situ deposition. Transport measurement indicates that Co films grow via Volmer-Weber type of growth on both the substrates. However, the islands coalesce at ~ 0.6 nm and at ~ 1.5 nm film thicknesses for smooth and rough substrate respectively. The evaluation of hysteresis loops with increasing Co layer thickness indicates ferromagnetic phase formation at lesser film thickness for smooth substrate compared to rough substrate. Azimuthal angle dependent MOKE measurement displays presence of weak UMA. Long-range stresses may be a possible cause of UMA, which might have occurred during the deposition. Increase in domain walls pinning or demagnetization factor in rough film is found responsible for the increased value of Hc. Therefore, the variation of Hc and UMA between two films are attributed to a combined effect of surface morphology and internal stresses. Compared to it, film deposited over periodic rippled pattern of comparable surface roughness is found to induce stronger UMA originating due to shape anisotropy of the sinusoidal wrinkled film topography. The results indicate that interface morphology has crucial bearings on growth, morphology, coercivity, magnetic anisotropy and provides an advancement to this field of research. Therefore, a proper characterization of interfacial roughness and its impact on different material's properties must be fully explored before its direct application in next generation devices.

## ACKNOWLEGMENT

Authors acknowledge Dr. V. Ganesan and Mohan Gangarade for AFM measurements and Dr. V.R. Reddy for XRR measurements.